\documentclass[prd,showpacs,preprintnumbers,amsmath,amssymb,twocolumn,
superscriptaddress,notitlepage]{revtex4-2}

\usepackage{cases}
\usepackage{amsmath}
\usepackage{amssymb}
\usepackage{amsfonts}
\usepackage{amssymb}
\usepackage{dcolumn}
\usepackage{bm}
\usepackage{epsfig}
\usepackage{bbm}
\usepackage{graphicx}
\usepackage{xcolor}
\usepackage{array}
\usepackage{subfigure}
\usepackage{hyperref}
\usepackage{mathrsfs}
\usepackage{verbatim}
\usepackage{float}
\usepackage{epstopdf}
\usepackage{color}
\usepackage{orcidlink}
\usepackage{comment}

\newcommand{\bra}[1]{\langle #1\vert}
\newcommand{\ket}[1]{\vert #1\rangle}
\newcommand{\be}{\begin{equation}}
\newcommand{\ee}{\end{equation}}
\newcommand{\ba}{\begin{eqnarray}}
\newcommand{\ea}{\end{eqnarray}}

\newcommand{\gsim}{\mathrel{\hbox{\rlap{\lower.55ex \hbox {$\sim$}}
			\kern-.3em \raise.4ex \hbox{$>$}}}}
\newcommand{\lsim}{\mathrel{\hbox{\rlap{\lower.55ex \hbox {$\sim$}}
			\kern-.3em \raise.4ex \hbox{$<$}}}}

\usepackage{ulem,fancyvrb}
\usepackage{xcolor}

\newcommand{\jl}[1]{\textcolor{orange} {jl: #1}}

\hypersetup{colorlinks=true,
	breaklinks=true,
	pdfstartview=Fit,
	linkcolor=blue,
	citecolor=blue,
	urlcolor=blue}

\begin{document}
\title{An Axial-Vector Leptophilic Fifth Force Sourced by Solar Neutrinos}

\author{Rundong Fang$ $} 
\thanks{These authors contributed equally to this work.}
\affiliation{School of Physics, Beihang University, Beijing 100083, China}
\affiliation{Center for High Energy Physics, Peking University, Beijing 100871, China}

\author{Ji-Heng Guo$ $}
\thanks{These authors contributed equally to this work.}
\affiliation{School of Physics, Beihang University, Beijing 100083, China}
 
\author{Jia Liu \orcidlink{0000-0001-7386-0253}}
\thanks{Contact author}
\email{jialiu@pku.edu.cn}
\affiliation{School of Physics and State Key Laboratory of Nuclear Physics and Technology, Peking University, Beijing 100871, China}
\affiliation{Center for High Energy Physics, Peking University, Beijing 100871, China}

\author{Xiao-Ping Wang \orcidlink{0000-0002-2258-7741} }
\thanks{Contact author}
\email{hcwangxiaoping@buaa.edu.cn}
\affiliation{School of Physics, Beihang University, Beijing 100083, China}
\affiliation{Beijing Key Laboratory of Advanced Nuclear Materials and Physics, Beihang University,
Beijing 100191, China}

\author{YanLi Zhao${}$}
\thanks{Contact author}
\email{sy2419177@buaa.edu.cn}
\affiliation{School of Physics, Beihang University, Beijing 100083, China}

\begin{abstract}
We investigate long-range, purely leptophilic axial--vector interactions mediated by a light gauge boson \(A'\) that couples to charged leptons and, by weak symmetry, to left-handed neutrinos. We analyze two realizations, a minimal effective model with muon-only couplings and an anomaly-free axial \(U(1)'\) with inter-generation cancellations. In both cases, the solar neutrino flux acts as an extended current that sources a macroscopic \(A'\) field at Earth, with spatial components aligned along the Sun--Earth direction. This field produces a distinctive signature in storage-ring measurements of the muon anomalous magnetic moment, \((g-2)_\mu\), namely a diurnal, sign-changing contribution that is positive during daytime and negative at night, superimposed on a time-independent positive offset. We obtain bounds \(g' \lesssim \mathcal{O}(10^{-19})\) in both model frameworks for a light, effectively massless mediator. For completeness, we map the solar-neutrino--sourced potential to electron spin-sensor experiments and find \(g' \lesssim \mathcal{O}(10^{-22})\) in the electron channel.

\end{abstract}

\maketitle

\section{Introduction}

Possible new and feeble interactions beyond the Standard Model (SM) and gravity—often dubbed “fifth forces”—span many orders of magnitude in mediator mass and interaction range~\cite{RevModPhys.97.025005}. A broad suite of precision measurements across nuclear, atomic, and condensed-matter platforms has been deployed to probe such forces among nucleons~\cite{Venema:1992zz, Glenday:2008zz, Vasilakis:2008yn, Su:2021dou, Tullney:2013wqa, Wei:2022mra, Zhang:2023qmu} and electrons~\cite{Ni:1999di, Chu:2015tha, Crescini:2020ykp, Vorobev:1988ug, Huang:2024esg}. In the nonrelativistic limit, exchanges of light bosons with scalar, pseudoscalar, vector, or axial-vector (AV) couplings generate a complete set of spin-independent and spin-dependent potentials~\cite{Dobrescu:2006au}. Within the lepton sector, the electron–electron potential from AV exchange is intrinsically spin-dependent and has been extensively constrained by torsion pendula, magnetometers, and solid-state spin sensors, which translate null results into tight bounds on the electron coupling~\cite{Heckel:2013ina,Hunter:2013hza,PhysRevLett.115.201801,PhysRevLett.125.201802}.

Most existing studies effectively target charged leptons, frequently assuming a vector coupling to nucleons so that macroscopic sources such as the Earth or the Sun generate a potential acting on electron spins. By contrast, $SU(2)_L$ symmetry suggests that left-handed neutrinos and their charged-lepton partners should share equal-magnitude couplings to new interactions, implying that neutrinos themselves can act as sources for a leptonic fifth-force potential. While long-range forces or non-standard interactions (NSI) in the neutrino sector have been widely examined, the emphasis has largely been on vector or scalar potentials sourced by electrons or nucleons~\cite{Gonzalez-Garcia:2006vic, Bandyopadhyay:2006uh, Gonzalez-Garcia:2008jtw, Samanta:2010zh, Wise:2018rnb, Smirnov:2019cae, Chauhan:2024qew, Coloma:2020gfv, Agarwalla:2024ylc, Mishra:2024riq, Fang:2024syu, Fang:2024wsg}. Axial-vector couplings of neutrinos have typically been discussed only in the short-range (heavy-mediator) NSI limit~\cite{Gehrlein:2024vwz}, leaving the long-range AV regime comparatively underexplored.

In this work, we develop a purely leptophilic AV–AV fifth-force framework mediated by a light $U(1)'$ gauge boson $A'$ that couples axially to both charged leptons and neutrinos. Because a purely axial $U(1)'$ is generically anomalous, we construct anomaly-free charge assignments via inter-generation cancellations, i.e., an axial analogue of the familiar $U(1)_{L_\mu-L_\tau}$ pattern. In this setup, the left-handed neutrino current is the operative source. For relativistic left-handed neutrinos, the AV coupling effectively reduces to a vector interaction with the chiral current, so an ambient neutrino flux plays the role of an electromagnetic-like source for the $A'$ field.

The solar neutrino flux provides the dominant, extended source for the long-range $A'$ potential at Earth. Atmospheric neutrinos are subdominant because of their lower flux, and the cosmic neutrino background yields a negligible net potential due to the presence of both neutrinos and antineutrinos. We derive the analytic form of the sourced $A'$ field and compute its value at Earth using a realistic solar flux profile.

The sourced $A'$ field has direct phenomenological consequences. If $A'$ couples in the $\mu-\tau$ pattern, the field acts on muon spins and modifies their precession in storage-ring experiments, entering the muon $g-2$ observable. This provides a sensitive probe of the axial muon coupling; we quantify the resulting limits and present constraints on the $U(1)'$ charges of $\mu$ and $\tau$. For the electron case, the same solar-neutrino–sourced potential affects electron-spin sensors. By mapping existing sensor results~\cite{Heckel:2008hw,Hunter:2013hza} onto our leptophilic AV framework, we find that the translated bounds are weaker by roughly one to two orders of magnitude than the current best electron limits derived using the polarized electrons inside Earth as source.

The remainder of the paper is organized as follows. Section~\ref{sec:model} introduces the anomaly-free UV setup. Section~\ref{sec:potential} derives the $A'$ potential sourced by neutrino fluxes. Section~\ref{sec:oscillation_probability} evaluates the fraction of each flavor of solar neutrinos from neutrino oscillation.
 Section~\ref{sec:gminus2} presents the implications for muon $g-2$ and the ensuing constraints, which are summarized for the $\mu-\tau$ pattern in Section~\ref{sec:mutau_constraints}. Section~\ref{sec:electron} treats the electron case and the translation to spin-sensor limits. We conclude in Section~\ref{sec:concl}.

\section{Axial-vector $\rm U(1)'$ model}
\label{sec:model}

We consider a $\rm U(1)'$ gauge boson $A'$ that couples to Standard Model (SM) fermions with a purely axial-vector structure~\cite{Ismail:2016tod}. In this work we focus on the lepton sector and assume that quarks are uncharged under $\rm U(1)'$. The general interaction Lagrangian is~\cite{Fang:2024wsg}
\begin{align}
\mathcal{L}_{\rm int}
= A'_{\mu}\!
\left( \boldsymbol{g}_L^{ij}\, \bar{L}_{i} \gamma^{\mu} L_{j}
+ \boldsymbol{g}_R^{ij}\, \bar{\ell}_R^{i} \gamma^{\mu} \ell_R^{j} \right),
\label{eq:5th-force-AV-lagrangian}
\end{align}
where $L=(\nu_L,\ell_L)$ is the left-handed $SU(2)_L$ lepton doublet, $\ell_R$ is the right-handed charged-lepton singlet, and $i,j=e,\mu,\tau$ are flavor indices.

Since the coupling is purely axial vector, and imposing that the $\rm U(1)'$ interactions are flavor diagonal~\footnote{With flavor-diagonal $U(1)'$ charges, the charge generator commutes with the diagonal charged-lepton Yukawa matrices, so the SM Higgs need not carry a $U(1)'$ charge. By contrast, flavor off-diagonal charges keep the gauge interaction gauge-invariant, but the Yukawa sector then requires additional structure (e.g., extra scalars) to keep gauge invariance~\cite{Ismail:2016tod}.}, we have
\begin{align}
\boldsymbol{g}_R^{ij} = -\,\boldsymbol{g}_L^{ij} \equiv g' Q_i\, \delta_{ij},
\end{align}
where $g'$ is the $\rm U(1)'$ gauge coupling and $Q_i$ is the charge of lepton generation $i$. 
The interaction Lagrangian then becomes
\begin{align}
\mathcal{L}_{\rm int}
= g' A'_\alpha \sum_{i=e,\mu,\tau} Q_i 
\left(
-\, \overline{\nu_{i,L}} \gamma^\alpha \gamma^5 \nu_{i,L}
+ \bar{\ell}_i \gamma^\alpha \gamma^5 \ell_i
\right).
\label{eq:anomaly_free_model}
\end{align}

Next we consider the anomaly-free conditions for the additional $\rm U(1)'$ gauge symmetry. The relevant triangle anomalies involve the following gauge structures~\cite{Ismail:2016tod}:
\begin{equation}
\begin{aligned}
&SU(2)_L^2 \times U(1)', \, SU(3)_c^2 \times U(1)', \,
U(1)_Y^2 \times U(1)',\\
&U(1)'^{\,3}, \,  U(1)_Y \times U(1)'^{\,2},  \,
[{\rm Gravity}]^2 \times U(1)',
\end{aligned}
\end{equation}
with the SM gauge group $SU(3)_c \times SU(2)_L \times U(1)_Y$. Requiring cancellation of each anomaly coefficient yields, in our setup,
\begin{equation}
\sum_{i=e,\mu,\tau} Q_i \;=\; 0,
\qquad
\sum_{i=e,\mu,\tau} Q_i^{3} \;=\; 0.
\label{eq:anomaly_condition}
\end{equation}
Nontrivial solutions to Eq.~\eqref{eq:anomaly_condition} take the form where one lepton has zero charge and the remaining two have equal magnitude and opposite sign. To evade constraints from laboratory spin-sensor searches that probe electron spins, we choose
\begin{align}
Q_e = 0, \qquad Q_\mu = -\,Q_\tau = 1,
\end{align}
without loss of generality. This charge pattern corresponds to the familiar $U(1)_{L_\mu-L_\tau}$ structure~\cite{He:1990pn,Foot:1990uf}. We therefore refer to our setup as the $L^{\rm AV}_\mu - L^{\rm AV}_\tau$ model. The relevant part of the Lagrangian describing the muon sector is then
\begin{align}
\mathcal{L}_{\mu}^{\rm AV}
= g' Q_{\mu} \, A'_\alpha
\left(
-\, \overline{\nu_{\mu,L}} \gamma^\alpha \gamma^5 \nu_{\mu, L}
+ \bar{\mu}\, \gamma^\alpha \gamma^5 \mu
\right),
\label{eq:muon_only_model}
\end{align}
where the $\mathrm{U(1)'}$ gauge boson $A'_\alpha$ couples exclusively to the muon generation,
which we denote it as $L^{\rm AV}_\mu$ model.
For clarity, we denote the flavor-dependent coupling as $g'_i \equiv g' Q_i$ ($i = e,\mu,\tau$).
This setup simplifies the phenomenology by isolating the muon and muon-neutrino interactions. Anomaly cancellation in this effective description can be restored by introducing additional heavy chiral fermions charged under $\mathrm{U(1)'}$, whose details are left unspecified.

\section{Axial-Vector Field Production from Solar Neutrinos}
\label{sec:potential}

In this section, we investigate how an axial-vector field $A^{\prime}$ can be generated by solar neutrinos. 
Solar neutrinos are exclusively electron neutrinos when produced inside Sun, and can oscillate into other flavor eigenstates (muon and tau) during their propagation.
% Solar neutrinos are primarily produced in the Sun via electron neutrino oscillations, which can naturally lead to the creation of muon neutrinos. 
These muon neutrinos can then interact with the new axial-vector gauge boson $A^{\prime}$, which mediates the interaction between the neutrino and the field. This interaction is governed by the coupling between the neutrinos and the axial-vector field, leading to a source term in the equation of motion for $A^{\prime}$. The resulting dynamics can be derived from the Euler-Lagrange equation for the axial-vector field:
\begin{equation}
	\partial_{\beta} F'^{\beta\alpha} + m^2_{A'} A'^{\alpha} = -g' \sum_{i=e,\mu,\tau} Q_i \bar\nu_i \gamma^\alpha \gamma^5 P_L \nu_i.
\label{EL-equation}
\end{equation}
For massive gauge field, taking the derivation to both side of the Eq.~\eqref{EL-equation} gives~\cite{Kikuchi:1972zf}
\be
m^2_{A} \partial_{\alpha}  A^{\alpha} =  -g' \partial_{\alpha} J^{\alpha},  
\ee
where $J^\alpha=\sum_{i=e,\mu,\tau} Q_i \bar\nu_i \gamma^\alpha \gamma^5 P_L \nu_i$ is the neutrino current. This shows that the axial-vector field is sourced by the neutrino current. The Lorentz gauge condition, $\partial_\alpha A^\alpha=0$, holds only when charge conservation is respected, i.e., when $\partial_\alpha J^\alpha=0$. However, for massless neutrinos, the current is always conserved. Since the energy of solar neutrinos is much larger than their mass, the approximation of massless neutrinos is well-justified, and the Lorentz gauge condition can be safely applied to the equation of motion for the field.

The neutrino current $J^\alpha$ is given by:
\begin{equation}
\bra{0}J^\alpha\ket{0} = \int \frac{d^3 \vec{p}_\nu}{2 E_{\nu}} \sum_{i, s} Q_i \rho^i_\nu(\vec{p}_\nu)
\bar{u}^s(p_\nu) \gamma^\alpha \gamma^5 P_L  u^{s}(p_\nu).
\end{equation}
where $\rho^i_\nu(\vec{p}_\nu)$ is the momentum distribution of the solar neutrino in the $i$-th flavor eigenstate, and $\bar{u}^s\left(p_\nu\right)$ and $u^s\left(p_\nu\right)$ are the spinor solutions for the neutrino field. For the spinor contraction $\bar{u}^s(p_\nu) \gamma^\alpha \gamma^5 P_L {u}^{s^{\prime}}(p_\nu)$, we have~\cite{Duda:2001hd}
\begin{equation}
\small{
\begin{aligned}
\bar{u}^s(p_\nu) \gamma^0 \gamma^5 P_L {u}^{s^{\prime}}(p_\nu) &=  \left(E_{\nu} - \vec{p}_{\nu} \cdot \vec{\sigma}_{\nu} \right)\delta^{s s^{\prime}},\\
\bar{u}^s(p_\nu) \gamma^i \gamma^5 P_L  {u}^{s^{\prime}}(p_\nu) &=  (p_{\nu}^i - m_{\nu} \sigma^i - \frac{\left(E_{\nu} - m_{\nu}\right) \vec{p}_{\nu}\cdot \vec{\sigma}_{\nu}}{|\vec{p}_{\nu}|^2}  p_{\nu}^i  ) \delta^{s s^{\prime}},
\end{aligned}}
\end{equation}
where $p_\nu$ ($E_{\nu} $, $\vec{p}_{\nu}$, $m_{\nu}$) is the four-momentum (energy, 3-momentum, mass) of the neutrino. In the relativistic limit, the current takes a much simpler form \begin{equation}
\bar{u}^s(p_\nu) \gamma^\alpha \gamma^5 P_L  {u}^{s^{\prime}}(p_\nu) = p_{\nu}^{\alpha}(1 - 2 h_\nu),
\end{equation}
where $h_\nu = \vec{p}_{\nu} \cdot \vec{\sigma}_{\nu}/ 2 |\vec{p}_{\nu}|$ is the helicity of the neutrino.

We now consider the calculation of the axial-vector field $A^{\prime}$ generated at the detector. We assume that the solar neutrino flux will not vary with time. Taking the relativistic limit and all of the solar neutrino to be left-handed, the solar neutrino generated axial-vector field $A'$ at the detector is given by
\be
A'^{\alpha} (\vec{x}_d)= \int d^3 x g' q_\nu^{\rm eff} (x)  { \Phi^{\alpha}(x) }\frac{1}{4 \pi \Delta r}  e^{-m_{A'} \Delta r},
\label{eq:axial-vector-potential}
\ee
where $\vec{x}_d$ is the position of the detector, and $\Delta r = |\vec{x}_d - \vec{x}|$ is the distance between the detector and the source, $q_\nu^{\rm eff} (x)$ is the effective charge for neutrino flux, defined as
\begin{align}
    q_\nu^{\rm eff} (x) & = \sum_{i= e,\mu,\tau} Q_i f_i(x), \\
   g' q_\nu^{\rm eff} (x) & = \sum_{i= e,\mu,\tau} g'_i f_i(x) .
\end{align}
where $f_{i}(x)$ is the fraction of each flavor of neutrino at location $x$, with $f_e(x) + f_\mu(x) + f_\tau(x) =1$. For solar neutrino, the fractions are equal to the oscillation probability from electron neutrino to the neutrino with corresponding flavor, which will be discussed in Sec.~\ref{sec:oscillation_probability}. The total neutrino flux $\Phi^{\alpha}(x)$ is
\begin{equation}
\Phi^{\alpha}(x) = n^{0}_{\nu} \left(\frac{1\, {\rm AU}}{\left|x\right|}\right)^2\frac{ p_\nu^{\alpha} }{ E_\nu },
\end{equation} 
with $n^{0}_{\nu} = 2.17\ {\rm cm}^{-3}$~\cite{Vitagliano:2019yzm} as the total solar neutrino density at Earth and AU being the astronomical unit.

Because the distribution of solar neutrinos is spherically symmetric, the direction of the axial-vector field $\vec{A}^{\prime}\left(\vec{x}_d\right)$ is aligned with the vector connecting the Sun and the detector. 
To simplify the calculation of the Sun's position relative to the detector, we neglect the Earth's radius and assume that Earth's orbit around the Sun is circular. In this simplified model, we define the position of the Sun in the lab frame based on the time of year and the orientation of the Earth's axis. Using the winter solstice as the time origin and the zenith of the laboratory as the $z$-direction, the Sun's position at time $t$ is given by the following matrix product, which describes the combined effects of the Earth's rotation and orbital motion around the Sun:

\begin{align}
\small
\vec{x}_{\odot} (t)=&
\begin{pmatrix}
	\sin \theta_{\rm L} & 0 & -\cos \theta_{\rm L}  \\
	0 & 1 & 0\\
	\cos \theta_{\rm L} & 0 & \sin \theta_{\rm L}
\end{pmatrix} 
\begin{pmatrix}
	\cos 2\pi\omega_d t & \sin 2\pi\omega_d t & 0 \\
	-\sin 2\pi\omega_d t & \cos 2\pi\omega_d t & 0\\
	0 & 0 & 1
\end{pmatrix} \nonumber\\ 
&
\times\begin{pmatrix}
	\cos \theta_{\rm OE} & 0 & \sin \theta_{\rm OE} \\
	0 & 1 & 0\\
	-\sin \theta_{\rm OE} & 0 & \cos \theta_{\rm OE}
\end{pmatrix} 
\begin{pmatrix}
	\cos 2\pi\omega_y t \\
	-\sin 2\pi\omega_y t \\
	0 
\end{pmatrix},
\end{align}
where $\omega_y$ ($\omega_d$) is the frequency of Earth's evolution around Sun (rotation),
$\theta_{\rm OE}$ is the obliquity of the ecliptic and $\theta_{\rm L}$ is the latitude of the laboratory.
Then we have the axial-vector field at the detector is:
\be
\vec{A}' (\vec{x}_d) = - |\vec{A}' (R)| \vec{x}_{\odot} (t)
\ee
where $R = |\vec{x}_d|$, and the minus sign means the direction from the Sun to the Earth.

\section{Solar Neutrino Oscillation Probability}
\label{sec:oscillation_probability}

In this section, we calculate the oscillation probability of solar neutrinos. 
Since most neutrinos are produced within the in the inner region of the Sun, we calculate the oscillation probability by treating different regions of the Sun separately. When neutrinos are propagating in the range $[0, 0.5 \,R_\odot ]$ where $R_\odot$ denotes the solar radii, we apply the adiabatic approximation to evaluate the revolution of the state~\cite{Mikheev:1986if, Bethe:1986ej}. This approximation hold when the density gradient is small compared with the neutrino oscillation length. Under this approximation, the eigenstates respect to the neutrino Hamiltonian 
\begin{equation}
\mathcal{H}_\nu 
\equiv  U \frac{m^2}{2 E_{\nu}}U^{\dagger} + {\rm diag}(\sqrt{2} G_F n_e, 0, 0) ,
\label{eq:fullH}
\end{equation}
will evolve independently of each other. Here
\( U \) stands for the PMNS matrix, \( m^2 = {\rm diag}(m^2_1, m^2_2, m^2_3) \) denotes the diagonal mass matrix, $n_e$ is the local electron density. 
We note that, in the Hamiltonian, we have ignored the contribution from the axial vector model, whose dominate contribution equals to $g' A'_0$~\cite{Brdar:2017kbt}. This is because its contribution is small in the region of interest. For example, for $m_{A'} = 0$, $E_\nu = 0.267$ MeV, and $g'= 10^{-18}$, we have $ 2 E_\nu g' A'_0 =  2.7 \times10^{-8}\ {\rm eV}^2$ at Earth, which is much smaller than the $m^2$ term. Meanwhile, when approaching to Sun, $A'_0$ will also decrease, leading to an even smaller effect.

Under the adiabatic approximation, the neutrino state at $0.5 \,R_\odot$ is given by~\cite{Maltoni:2015kca}
\begin{equation}
\ket{\nu(0.5 \,R_\odot)} = \sum_j U^{m*}_{e j}(n_e(r_0)) e^{-i \phi_j} \ket{\nu_j},
\label{eq:neutrino_state_half_sun}
\end{equation}
where $\ket{\nu_j}$ are the mass eigenstates, $r_0$ represent the neutrino production position, $U^m$ is the unitary matrix that diagonalize the total Hamiltonian at $r_0$. The phase is given by the integration of the eigenvalues of the Hamiltonian along the neutrino trajectory 
\begin{equation}
\phi_j = \int \mathcal{H}^j_\nu (z) d z.
\end{equation}

Outside the inner region $0.5\, R_\odot$, the neutrino propagation can be treated as vacuum oscillation, since the electron density in this region is much lower than that near the solar core~\cite{Grevesse:1998bj, Asplund:2009fu, Vinyoles:2016djt} and the energy of most neutrinos are also small. Therefore, the oscillation probability is given by the standard formula of the vacuum oscillation~\cite{Xing:2011zza}
\begin{equation}
\begin{aligned}
P_{e\to \alpha} = &\sum_{j} \left| U^*_{ej} U_{\alpha j} e^{- i \frac{m^2_j L}{2 E}}\right|^2\\
= &\delta_{e \alpha} - 4 \sum_{i<j} {\rm Re} \left( U^*_{ej} U_{\alpha j} U_{ei} U^*_{\alpha i}\right)\sin^2( \frac{\Delta m_{ij}^{2} L}{4 E_\nu})\\
&+ 2 \sum_{i<j} {\rm Im} \left( U^*_{ej} U_{\alpha j} U_{ei} U^*_{\alpha i}\right)\sin( \frac{\Delta m_{ij}^{2} L}{2 E_\nu}).
\label{P-outR}
\end{aligned}
\end{equation}
But before leaving the dense solar core $\left(r<0.5 R_{\odot}\right)$, neutrinos experience significant matter effects that modify their effective mixing. To incorporate these effects, we replace $U_{e j}^*$ with $U_{e j}^{m *}\left(n_e\left(r_0\right)\right) e^{-i \phi_j}$, where $U^m\left(n_e\left(r_0\right)\right)$ is the mixing matrix in matter evaluated at the production point $r_0$, and the phase factor $e^{-i \phi_j}$ represents the adiabatic evolution of each mass eigenstate as the neutrino propagates outward through the inner region.

The oscillation probability depends on the neutrino energy $E_\nu$ and propagation length $L$. When $L$ is much larger than the oscillation length, or when neutrinos have a small but finite energy spread $\delta E_\nu$, the rapidly oscillating interference terms tend to average out. This can be seen by expanding the oscillation phase:
\begin{equation}
\frac{\Delta m_{ij}^{2} L}{2 (E_\nu + \delta E_\nu)} \approx
\frac{\Delta m_{ij}^{2} L}{2 E_\nu}-
\frac{\Delta m_{ij}^{2} L}{2 E_\nu} \frac{\delta E_\nu}{E_\nu}.
\end{equation}
Averaging the probability over a small range of $\delta E_\nu$ causes the linear $\sin$ terms in Eq.~(\ref{P-outR}) to vanish, since the rapid phase variations wash out coherent interference.
As a result, the dependence on $L$ and $E_\nu$ disappears, and the long-distance oscillation probability takes the incoherent form~\cite{Maltoni:2015kca}:
\be
P_{e \to \alpha} = \sum_{j=1}^{3} \left|U^m_{e j} \left(n_e(r_0)\right)\right|^2 \left|U_{\alpha j}\right|^2.
\label{eq:solar_probability}
\ee

For solar neutrinos, the dominate source comes from the proton-proton (pp) chain reaction~\cite{Grevesse:1998bj, Asplund:2009fu, Vinyoles:2016djt}, with an average energy of 0.267 MeV and a maximum energy of 0.423 MeV~\cite{Vitagliano:2019yzm}. For neutrinos with energy $\bar{E}_{\mathrm{pp}}=$ 0.267 MeV , the corresponding oscillation lengths are:
\begin{align}
\frac{4\pi \bar{E}_{\rm pp}}{\Delta m^2_{21}}\approx 8.8~ {\rm km}, ~~~
\frac{4\pi \bar{E}_{\rm pp}}{\Delta m^2_{31}}\approx 0.27~ {\rm km},
\end{align}
which are much smaller than the Solar radius $R_\odot \approx 6.9 \times 10^5$ km. Therefore, we can safely employ Eq.~\eqref{eq:solar_probability} to calculate the oscillation probability outside the Sun. 

\begin{figure}[htbp]
\begin{centering}
\includegraphics[width=0.5 \textwidth]{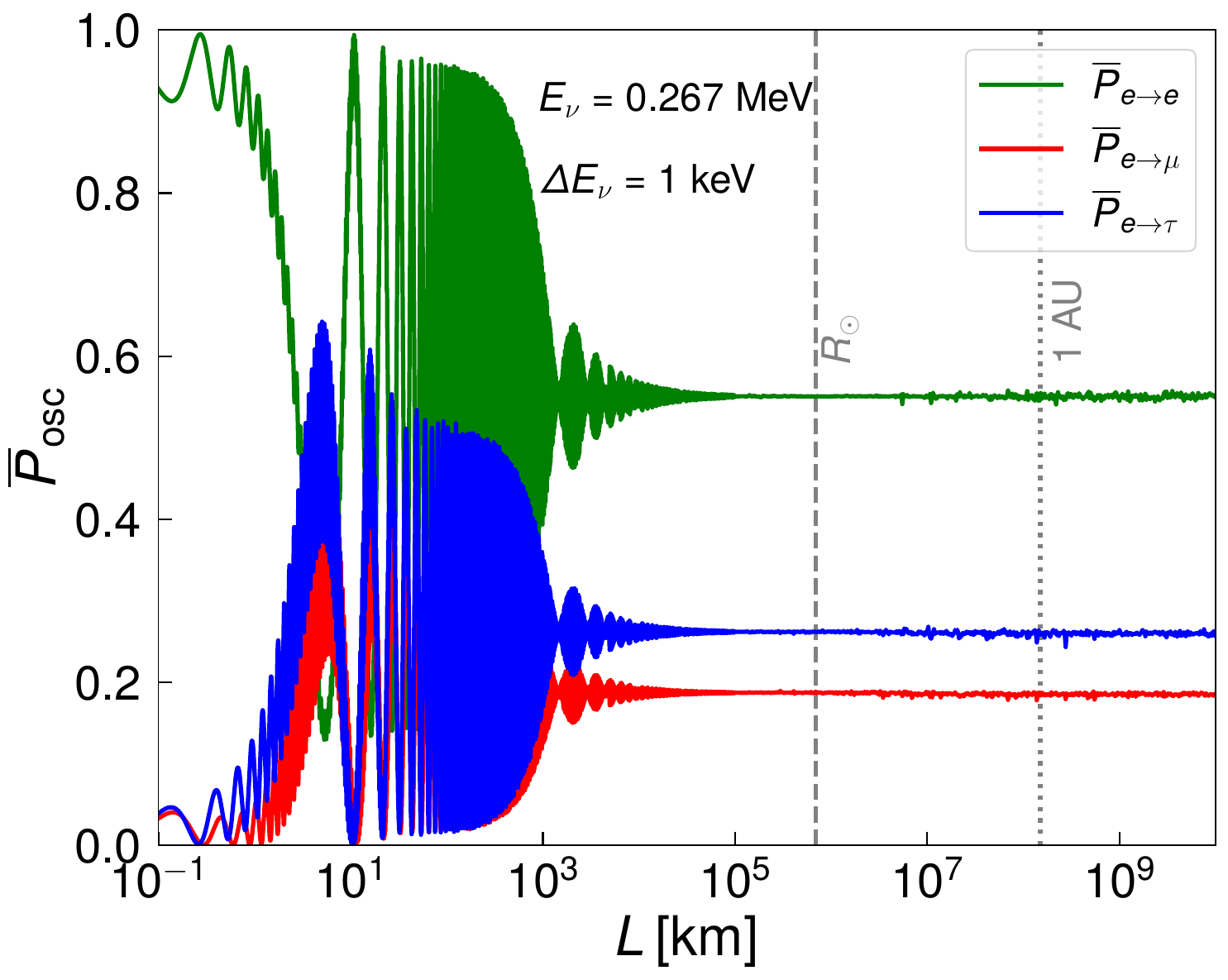}
\caption{The averaged oscillation probability for solar neutrinos with different propagation length. The green (red, and blue) solid line represents the probability that electron neutrino oscillate to electron (muon, and tau) neutrino.
The integration region have a center value at $E_\nu = \bar{E}_{\mathrm{pp}}=$ 0.267 MeV together with a width $\Delta E_\nu = 1$ keV in both side. The gray dashed (dotted) line represents the radii of Sun (Earth obit). 
}
\label{fig:averaged_probability_to_distance} 
\end{centering}
\end{figure}

A numerical calculation has also been performed as a cross-check to show that, for long enough propagation length, the oscillation probability in Eq.~\eqref{P-outR} can be well estimated by Eq.~\eqref{eq:solar_probability} after energy averaging. We define the energy-averaged oscillation probability as
\begin{equation}
\bar{P}_{e\to\alpha} (L) = \frac{1}{2 \Delta E_\nu} \int^{E_\nu+ \Delta E_\nu}_ {E_\nu- \Delta E_\nu} P_{e\to\alpha} (E, L) d E.
\end{equation}
In Fig.~\ref{fig:averaged_probability_to_distance}, we show the averaged oscillation probability $\bar{P}_{e\to\alpha}$ as a function of the propagation distance, where we take $E_\nu = \bar{E}_{\mathrm{pp}}=$ 0.267 MeV and $\Delta E_\nu = 1$ keV.
The electron density is taken from GS98~\cite{Grevesse:1998bj} data and the oscillation parameters are based on the best-fit results from Super-Kamiokande and SNO~\cite{Super-Kamiokande:2023jbt} for $\Delta m^2_{21}$, $\sin^2 \theta_{12}$ and the PDG values for other parameters~\cite{ParticleDataGroup:2024cfk}.
At $r = 0.05 R_\odot$, we have
\begin{equation}
\begin{aligned}
\left(U \frac{m^2}{2 E_{\nu}}U^{\dagger}\right)_{11} &= 1.35\times 10^{-10}\ {\rm eV}\\
\sqrt{2} G_F n_e (0.05 \,R_\odot) &= 6.82 \times 10^{-12}\ {\rm eV},
\label{eq:hamiltonian_compare}
\end{aligned}
\end{equation}
which shows that the matter potential term is much smaller than the vacuum term in the Hamiltonian of Eq.~\eqref{eq:fullH}. Hence, we neglect the matter term in the following numerical treatment. For propagation distances shorter than $\mathcal{O}(10^4)$~km, the averaged oscillation probabilities exhibit strong dependence on $L$. However, beyond $\sim10^5$~km, the probabilities converge to constant asymptotic values, namely
\begin{align}
\bar{P}_{e\to e} = 0.55 , ~~ \bar{P}_{e\to \mu} = 0.19 , ~~~\bar{P}_{e\to \tau} = 0.26,
\end{align}
which are the same as the results given by Eq.~\eqref{eq:solar_probability}. The small residual fluctuations arise from numerical inaccuracies due to integrating rapidly oscillating functions and diminish with higher integration precision. This behavior is fully consistent with analytical expectations. We therefore conclude that, for neutrinos propagating outside the Sun, the oscillation probabilities can be well described by Eq.~\eqref{eq:solar_probability}, which are effectively independent of the propagation length.

\begin{figure}[htbp]
\begin{centering}
\includegraphics[width=0.5 \textwidth]{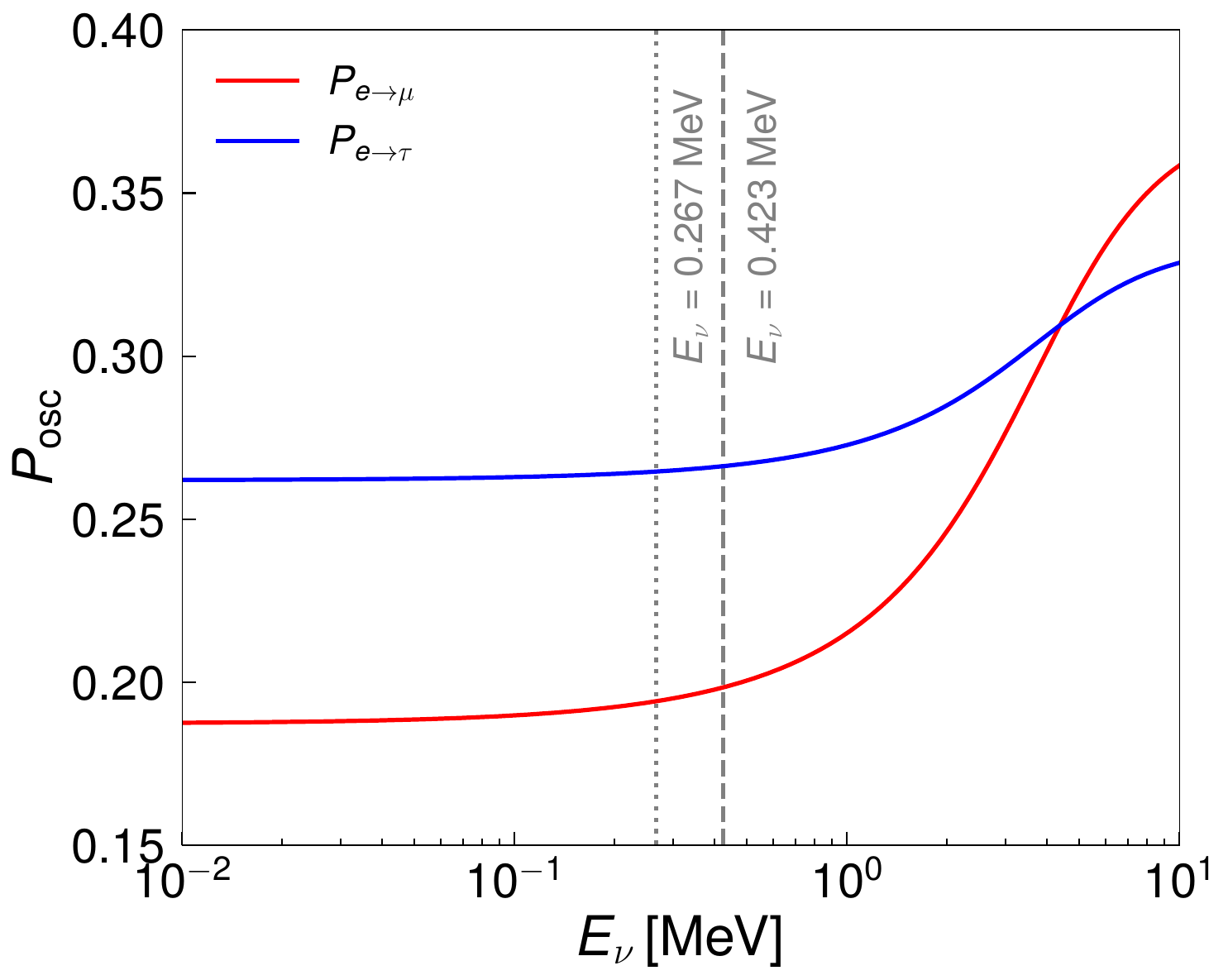}
\caption{The oscillation probability for solar neutrinos the oscillation probability is given by Eq.~\eqref{eq:solar_probability}, therefore it do not have any information of $r$. However, as such equation is usually used to calculate the oscillation probability at Earth, maybe we can take $r = 1$ AU as a function of neutrino energy. The red (blue) solid line represents the probability that electron neutrino oscillate to muon (tau) neutrino. The gray dashed (dotted) line represents the maximal (averaged) energy of pp chain neutrino.
}
\label{fig:oscillation_probability} 
\end{centering}
\end{figure}

For the oscillation probability in Eq.~\eqref{eq:solar_probability}, although the fast oscillating terms have been averaged out, it will still depend on the neutrino energy due to the $U^m(n_e(r_0))$ depends on the neutrino energy.  Fig~\ref{fig:oscillation_probability} shows the oscillation probability given by Eq.~\eqref{eq:solar_probability} as a function of neutrino energy, where the neutrino is assumed to be produced at $0.05 R_\odot$. At low energies, as indicated by Eq.~\eqref{eq:hamiltonian_compare}, the vacuum term $U m^2 U^{\dagger}/(2E_\nu)$ dominates over the matter potential $\sqrt{2} G_F n_e$, so the oscillations are well approximated by the vacuum case and the probabilities remain nearly constant.
As $E_\nu$ increases, the vacuum term decreases as $1/E_\nu$ while the matter potential stays constant, leading to stronger matter effects and noticeable variation in the oscillation probabilities.
However, since the maximum pp-chain neutrino energy is only $0.423~\mathrm{MeV}$, these neutrinos lie in the low-energy regime where the vacuum approximation holds, as seen in Fig.~\ref{fig:oscillation_probability}.

Consequently, in the following calculations, for neutrino outside Sun, we fix the oscillation probabilities as $P_{e \to e} = 0.55$, $P_{e \to \mu} = 0.19$ and $P_{e \to \tau} = 0.26$. 
The contribution from neutrinos inside Sun will be neglected as the their total number is much smaller than that outside Sun.

\section{axial-vector effect on muon $g-2$}
\label{sec:gminus2}

In this section, we investigate the effect of the axial vector boson $A^{\prime}$ on the muon $g-2$ measurements, focusing on how the muon's axial vector coupling $g'_{\mu}$ influences the observed results. In the experiment, muons are trapped in a magnetic field, where their spin precesses due to the interaction with both the external field and potential new physics contributions, such as the axial vector interaction with the $A^{\prime}$ boson. The interaction Hamiltonian for the system is then given by
\begin{equation}
H_{\rm int} = g' A'_{\alpha} J^{\alpha}_\mu,
\end{equation}
where $J_\mu$ is the muon current, defined as:
\begin{equation}
J^{\alpha}_\mu = \int \frac{d^3 p_{\mu}}{2 E_\mu} \bar{u}^s(p_\mu) \gamma^\alpha \gamma^5 {u}^{s^{\prime}}(p_\mu) \rho(\vec{p}_{\mu}),
\end{equation}
with $\rho(\vec{p}_{\mu})$ being the muon momentum distribution. In the laboratory frame, the momentum distribution is sharply peaked, and can be approximated by a Dirac delta function. Under this assumption, the axial current components become
\begin{equation}
\begin{aligned}
\overline{u}(p_\mu)  \gamma^0 \gamma^5 u(p_\mu) &= 2 \vec{p}_\mu \cdot \vec{\sigma}_\mu,\\
\overline{u}(p_\mu)  \gamma^i \gamma^5 u(p_\mu) &=  2m_\mu \sigma_\mu^i + 2 (E_\mu-m_\mu) \frac{\vec{p}_\mu \cdot \vec{\sigma}_\mu }{|\vec{p}_\mu|^2} p_\mu^i ,
\end{aligned}
\end{equation}
where $\vec{\sigma}_\mu$ denotes the muon spin vector. The momentum and spin are
\begin{equation}
\begin{aligned}
\vec{p}_\mu &=  (-P_\mu \sin (2\pi \omega_c t_l), P_\mu \cos (2\pi \omega_c t_l) ,0 ),\\
\vec{\sigma}_\mu &= (-\sigma_{xy} \sin (2\pi \omega_s t_l + \phi_0), \sigma_{xy} \cos (2\pi \omega_s t_l + \phi_0), \sigma_z),
\end{aligned}
\end{equation}
where $P_\mu$ is the total momentum of the muon in the lab frame, $\sigma_{xy}$ is the spin projection in the x-y plane, $\sigma_z$ is the spin of muon in the z-direction,
$t_l$ is the time in the lab frame,  $\omega_c$ ($\omega_s$) is the cyclotron (spin precession) frequency of muon in the lab frame, $\phi_0$ is an initial phase, which is taken as $\phi_0 = 0$.

To facilitate the calculation, we do a linear transformation as follows, 
\begin{equation}
H_{\rm int} = g' A'_{\alpha}  \Lambda^{\alpha\xi} \left(\Lambda^{-1}\right)_{\xi\alpha} J^{\alpha}_\mu,
\end{equation}
where the transformation matrix $\Lambda^{-1}$ is given by:
\begin{equation}
\Lambda^{-1} = 
\begin{pmatrix}
	\gamma_{\mu} & 0 & -\gamma_{\mu}\beta_{\mu}& 0 \\
	0 & 1 & 0 & 0\\
	-\gamma_{\mu}\beta_{\mu} & 0 & \gamma_{\mu} & 0\\
	0 & 0 & 0 & 1
\end{pmatrix} 
\begin{pmatrix}
	1&0 &0 &0\\
	0& c & s & 0 \\
	0&-s & c & 0\\
	0&0 & 0 & 1
\end{pmatrix},
\end{equation}
with $c \equiv  \cos 2\pi\omega_c t_l$, $s \equiv \sin 2\pi\omega_c t_l$, $\gamma_{\mu} = E_\mu/m_\mu$ and $\gamma_{\mu}\beta_{\mu} = P_\mu/m_\mu$.
This transformation brings the muon momentum to $(m_\mu , 0, 0, 0)$ and we have 
\begin{equation}
\left(\Lambda^{-1}\right)_{\xi\alpha} J^{\alpha}_\mu = (0, -\sigma_{xy} \sin \left(2\pi \omega_a  t_l \right), \sigma_{xy} \cos \left(2\pi \omega_a  t_l \right), \sigma_z),
\end{equation}
where $\omega_a = \omega_s -\omega_c$ is the anomalous precession frequency. 

Here we note that the transformation $\Lambda^{-1}$  also acts on the integration in $J_\mu$, which means that the muon momentum in the integral is transformed accordingly after applying $\Lambda^{-1}$. This transformation is not a Lorentz transformation, because a true Lorentz transformation is time independent, while $\Lambda$ explicitly depends on time. Nevertheless, for convenience, we refer to the transformed quantities as those in the rotating muon rest frame (RMRF). In this frame, the axial-vector potential is defined as
\begin{equation}
A'^{\xi}_{\rm RMRF} = A'_{\alpha}  \Lambda^{\alpha\xi}.
\end{equation}
where $A'_{\alpha} = g_{\alpha\beta} A'^{\beta} = (A'^0 , - \vec{A}')$.

The muon spin evolution obeys
\be
\frac{d\vec{S}}{d t} = \vec{\omega} \times \vec{S} \,, \quad
\frac{d \hat{S}_i }{d t} = i\left[ H, \hat{S}_i \right],
\ee
leading to a shift in the precession frequency,
\begin{equation}
\delta \vec{\omega} = 2 g' \vec{A}'_{\rm RMRF}.
\end{equation}
Therefore, the precession frequency shift is
\be
\Delta \omega = \sqrt{(\delta \vec\omega + \vec \omega)^2} - |\vec \omega|\approx \frac{|\delta \vec\omega|^2}{2 |\vec \omega|} + \delta \omega_z,
\label{eq:5th-deltaomega}
\ee
where $|\vec \omega| = \gamma_{\mu} \omega_a \equiv \gamma_{\mu} (\omega_c -\omega_s)$. Since the muon cyclotron period is much shorter than the solar modulation period, the Sun can be treated as a static background during one revolution. Averaging over a single cyclotron period yields
\be
\overline{\Delta \omega}(t) = \omega_c \int^{\frac{1}{\omega_c}}_0 \Delta \omega (t_l, t) d t_l.
\ee
where $t$ denotes the slow timescale associated with the Sun’s motion, uncorrelated with the lab-frame time $t_l$. The resulting averaged components are
\begin{widetext}
\begin{equation}
\overline{\delta \omega}_z(t) = 2 g'|\vec{A'}| \left(- \sin\theta_{\rm OE} \sin\theta_{\rm L} \cos 2 \pi \omega_y t + \cos\theta_{\rm L} (\cos\theta_{\rm OE} \cos 2 \pi \omega_y t  \cos 2 \pi \omega_d t  - \sin 2 \pi \omega_y t  \sin 2 \pi \omega_d t) \right),
\end{equation}

\begin{equation}
\begin{aligned}
\overline{|\delta \vec\omega|^2} (t)= 
&\frac{g^{\prime 2}}{4} \left( 16 A'^2_0 \gamma_\mu^2\beta_\mu^2+4 |\vec{A}'|^2 \left(-\gamma_\mu^2\beta_\mu^2 \sin \theta _{\rm OE} \sin 2 \theta _{\rm L}
\sin 2 \pi  \omega _d t \sin 4 \pi  \omega _y t \right. \right.\\
&\left. \left.+2 \cos ^2\theta _{\rm L} \left( (\gamma_\mu^2+1) \sin ^2\theta _{\rm OE} \cos ^2 2 \pi  \omega _y t+2 \sin ^2 2 \pi  \omega _d t \sin ^2 2 \pi  \omega _y t \right) \right. \right.\\
 &\left. \left.+2 (\gamma_\mu^2+1) \sin ^2 2 \pi  \omega _y t (\sin ^2 \theta _{\rm L} \sin^2 2 \pi  \omega _d t +\cos^2 2 \pi  \omega _d t ) \right. \right.\\
 &\left. \left. +\cos^2 2 \pi  \omega _y t  (\gamma_\mu^2\beta_\mu^2 \sin 2 \theta _{\rm OE} \sin 2 \theta _{\rm L} \cos 2 \pi  \omega _d t 
+4 \sin ^2\theta _{\rm OE} \sin ^2\theta _{\rm L})\right) \right.\\
 &  \left.+ |\vec{A}'|^2 \cos ^2\theta _{\rm OE} \cos ^2 2 \pi  \omega _y t (-4 \gamma_\mu^2\beta_\mu^2 \cos ^2 \theta _{\rm L} \cos 4 \pi  \omega _d t \right.
 -2 \gamma_\mu^2\beta_\mu^2 \cos 2 \theta _{\rm L}+6 \gamma_\mu^2+10)\\
 & \left.+4 |\vec{A}'|^2 \gamma_\mu^2\beta_\mu^2 \cos \theta _{\rm OE} \cos ^2\theta _{\rm L} \sin 4 \pi  \omega _d t \sin 4 \pi  \omega _y t \right).
\end{aligned}
\end{equation}
\end{widetext}

Finally, boosting back to the laboratory frame gives
\begin{equation}
\overline{\Delta \omega}_{\rm lab}(t)
= \frac{\overline{\Delta \omega}(t)}{\gamma_\mu}.
\end{equation}
We use the most recent experimental and theoretical values of the muon anomalous magnetic moment, $a_\mu^{\rm exp}$ and $a_\mu^{\rm SM}$, reported in Refs.~\cite{Aliberti:2025beg, Muong-2:2025xyk}, to constrain the parameter space.

\section{Results}
\label{sec:mutau_constraints}

\begin{figure}[htbp]
\begin{centering}
\includegraphics[width=0.5 \textwidth]{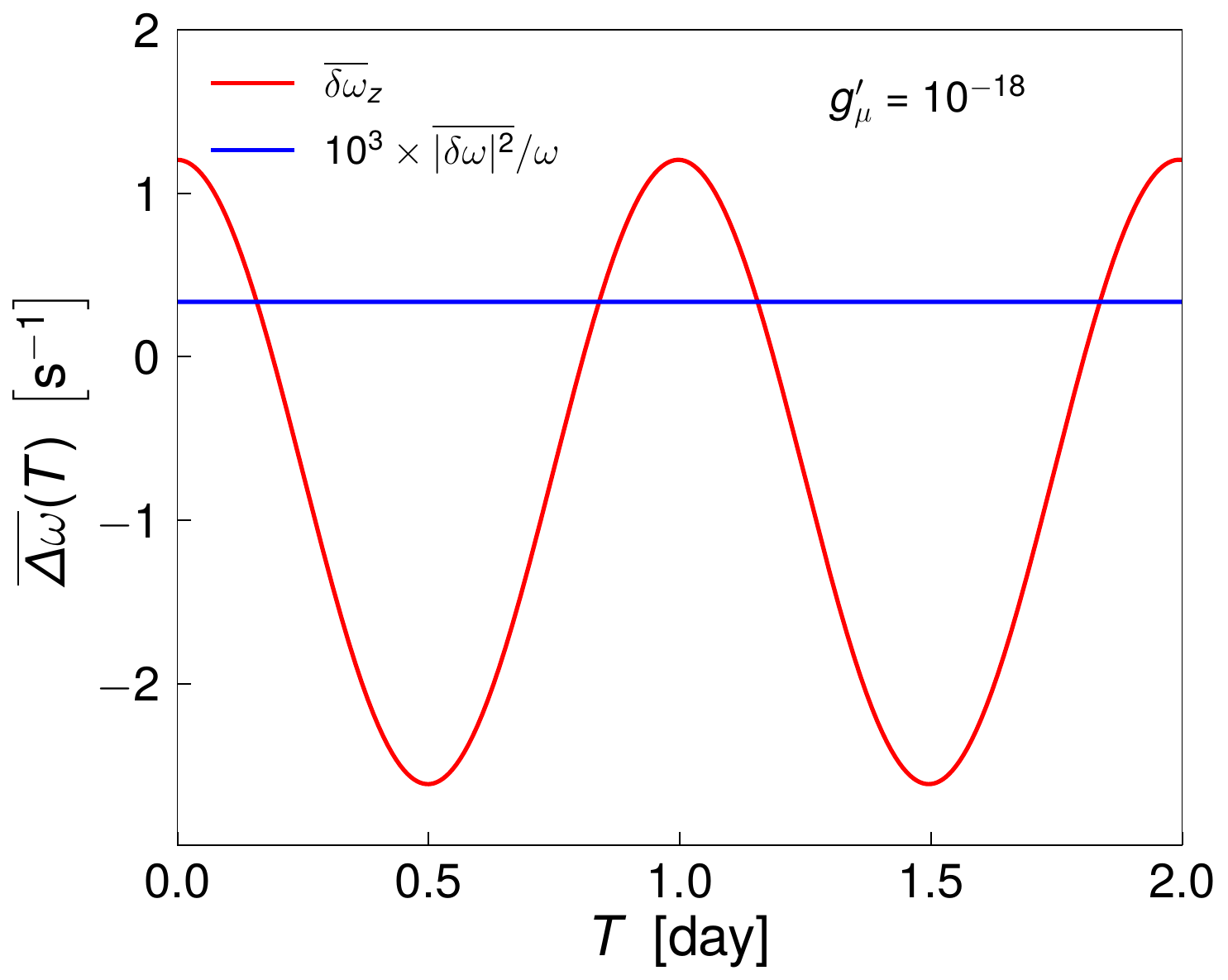}
\caption{The variation of the frequency respect to times in $L^{\rm AV}_\mu$ model. The red (blue) line represents the variation from $\delta \omega_z$ (${|\delta \vec\omega|^2}/{2 |\vec \omega|}$). Here we take $m_{A'} = 0$ and $g'_{\mu} = 10^{-18}$.
}
\label{fig:delta_omega} 
\end{centering}
\end{figure}

After including the axial-vector effect on the muon $g-2$, we can derive experimental constraints on the model parameters. We first examine the time-dependent precession frequency shift in the $L_\mu^{\rm AV}$ model, as shown in Fig.~\ref{fig:delta_omega}. The contribution from $\delta \omega_z$ exhibits a pronounced day-night asymmetry because it aligns with the solar direction; $\delta \omega_z$ is proportional to the cosine of the Sun's zenith angle. Using the winter solstice as the reference time in the laboratory frame, $\delta \omega_z$ remains negative for a longer period, with its minimum exceeding its maximum in magnitude. This pattern reverses at the summer solstice.

In contrast, the contribution from $|\delta \vec{\omega}|^2/\left(2|\vec{\omega}|\right)$ remains nearly constant over time, as $A_0^{\prime}$ is one order of magnitude larger than $\left|\vec{A}^{\prime}\right|$ in massless limit. This term is dominated by the time-independent factor $4 g^{\prime 2} A_0^2 \gamma_\mu^2 \beta_\mu^2$. Moreover, while $\delta \omega_z \propto g'|\vec{A'}|$, the term $|\delta \vec{\omega}|^2/\left(2|\vec{\omega}|\right)$ scales as $g^{\prime 2} A_0^2$. Consequently, for smaller values of $g'$, the contribution from $\delta \omega_z$ becomes more significant.
As shown in Fig.~\ref{fig:delta_omega}, when $g' = 10^{-18}$, the maximum of $\delta \omega_z$ is about three orders of magnitude larger than ${|\delta \vec\omega|^2}/{\left(2 |\vec \omega|\right)}$.

When setting constraints on the coupling strength of the proposed time-dependent new physics phenomenon from the Fermilab muon $g-2$ experiment, a precise comparison would require knowledge of the exact timing of data acquisition runs throughout the year. Since such detailed scheduling information is not publicly available, we adopt a conservative simplification: the theoretically predicted precession frequency shift, $\Delta \omega(t)$, is averaged over an entire year to obtain its mean value, $\left \langle \overline{\Delta \omega}(t) \right \rangle$. This average signal is then directly compared to the experimental uncertainty $\Delta a_\mu / a_\mu$ through the condition
\begin{equation}
\frac{\left \langle \overline{\Delta \omega}(t) \right \rangle  }{ \omega_a} < \frac{3 \Delta a_\mu }{ a_\mu}.
\label{eq:constraint}
\end{equation}
Form the latest result of muon g-2 experiment, we take 
$\Delta a_\mu / a_\mu = 1.27 \times 10^{-7}$ (127 ppb)~\cite{Muong-2:2025xyk}.
This annual-averaging procedure provides a robust and conservative estimate. Because the experimental data was collected over multiple years (2018-2024)~\cite{Foster:2023kzz}, the analysis inherently integrates over the potential temporal variation.  Consequently, our approach may slightly underestimate the experiment’s actual sensitivity to a time-varying effect, but it ensures that the resulting exclusion limits are both reliable and conservative.

Since the annual average of the ${\delta \omega}_z$ vanish, as a conserved estimation, we consider only the contribution from ${|\delta \vec\omega|^2}/{2 |\vec \omega|}$. This yields constraints for $m_{A'} = 0$ as 
\begin{align}
&g' < 4.8 \times 10^{-18} ~(L^{\rm AV}_\mu ~\rm case),\\
&g' < 8.0 \times 10^{-18} ~ (L^{\rm AV}_\mu - L^{\rm AV}_\tau ~ \rm case).
\end{align}

\begin{figure}[htbp]
\begin{centering}
\includegraphics[width=0.5 \textwidth]{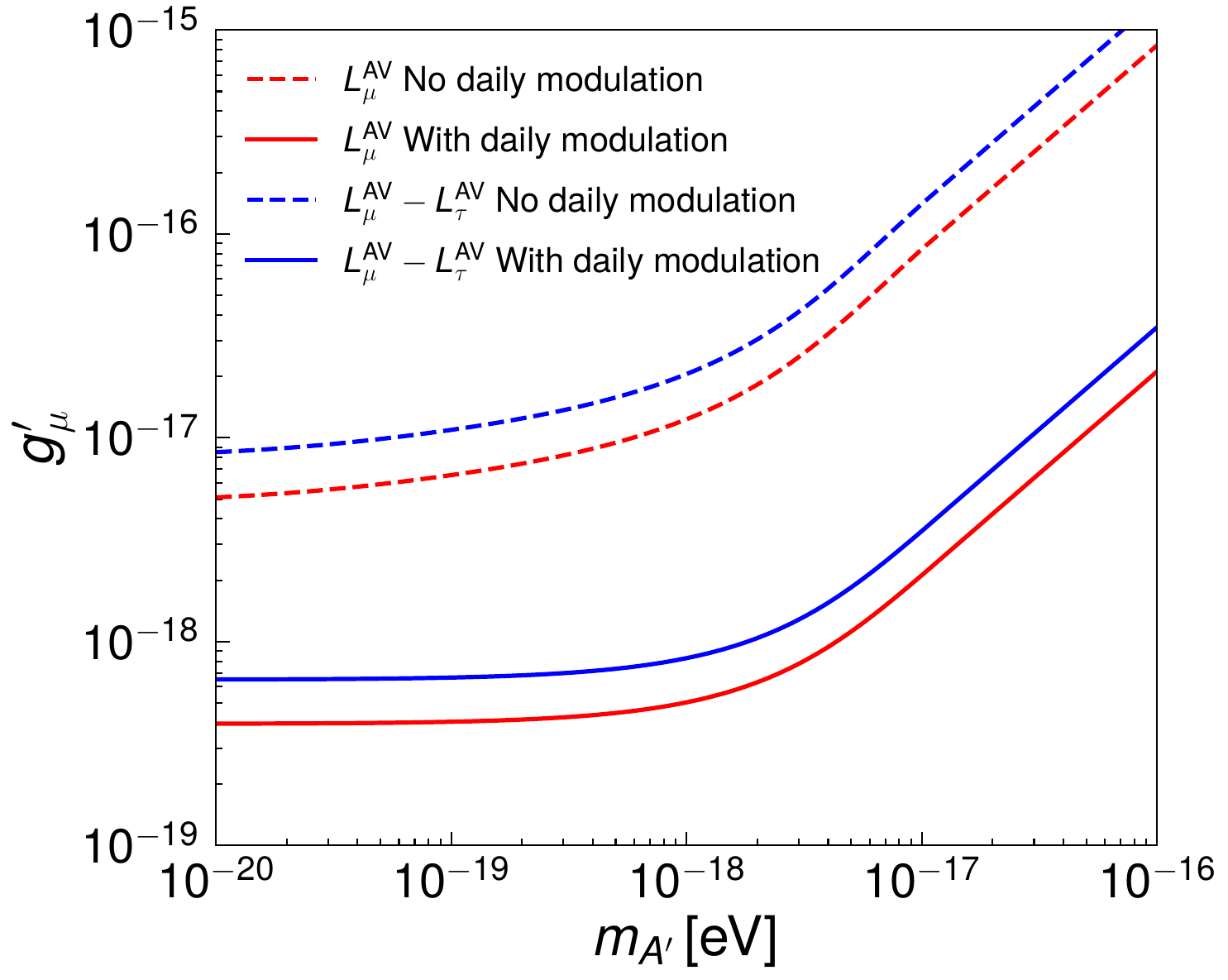}
\caption{Constraint to the $L^{\rm AV}_\mu$ model (red) and $L^{\rm AV}_\mu - L^{\rm AV}_\tau$ model from muon $g-2$ experiment~\cite{Muong-2:2025xyk}. The solid (dashed) line represent the constraint (not) considering the daily modulation.
}
\label{fig:constraint}
\end{centering}
\end{figure}

For a massive $A'$, the constraints remain similar to the massless case when $m_{A'} < R^{-1} \sim 10^{-18}~\mathrm{eV}$, where $R = 1$ AU is the Sun-Earth distance. However, in the high-mass limit where $m_{A'} \gtrsim 10^{-17}~\mathrm{eV}$, the potential scales as $(A'_0, \vec{A}') \propto m^{-2}_{A'}$ (see Appendix for a detailed discussion). This scaling leads to the constraints:
\begin{align}
&g' < 8.4 \times 10^{-17}\times \frac{ m_{A'}}{10^{-17}\  {\rm eV}} ~(L^{\rm AV}_\mu ~\rm case),\\
&g' < 1.4 \times 10^{-16}\times   \frac{ m_{A'}}{10^{-17}\  {\rm eV}} ~ (L^{\rm AV}_\mu - L^{\rm AV}_\tau ~ \rm case).
\end{align}
A significantly stronger constraint can be derived by exploiting the day-night asymmetry of $\delta \omega_z$. We estimate this effect by separately averaging the positive contributions $\delta \omega_z^{+}$ during the day and the negative contributions $\delta \omega_z^{-}$ during the night over a full year. Replacing $\langle\overline{\Delta \omega}(t)\rangle$ in Eq.~\eqref{eq:constraint} with the asymmetry $\left\langle\delta \omega_z^{+}\right\rangle-\left\langle\delta \omega_z^{-}\right\rangle$, we obtain for $m_{A'}=0$
\begin{align}
&g' < 4.0 \times 10^{-19}  ~(L^{\rm AV}_\mu ~\rm case),\\
&g' < 6.5 \times 10^{-19} ~ (L^{\rm AV}_\mu - L^{\rm AV}_\tau ~ \rm case).
\end{align}
For $m_{A'} \gtrsim 10^{-17}\ {\rm eV}$, we have 
\begin{align}
&g' < 2.1 \times 10^{-18}\times \frac{ m_{A'}}{10^{-17}\  {\rm eV}} ~(L^{\rm AV}_\mu ~\rm case),\\
&g' < 3.5 \times 10^{-18}\times   \frac{ m_{A'}}{10^{-17}\  {\rm eV}} ~ (L^{\rm AV}_\mu - L^{\rm AV}_\tau ~ \rm case).
\end{align}
We show the constraints in Fig.~\ref{fig:constraint}. The constraints from the daily modulation are approximately one order of magnitude stronger than those from the simple time average. This enhancement arises because the day-night asymmetry is proportional to $g'^2$, whereas the other contribution is proportional to $g'^4$. We note that our analysis, which relies on an annual average, does not incorporate the detailed time-dependent waveform of the signal. A future analysis using the full signal waveform could potentially yield even stronger constraints.

Other potential constraints for these models could arise from binary pulsar systems~\cite{Dror:2019uea, Liu:2025zuz} and rare Z decays~\cite{Dror:2017ehi, Dror:2017nsg}. Neutron stars, which contain a significant number of muons due to the high Fermi energy of their electrons, could radiate bosons coupled to muons, providing a means to constrain these couplings. For vector or scalar couplings, such constraints can reach down to $10^{-20}-10^{-21}$ in the massless limit~\cite{Dror:2019uea, Liu:2025zuz}. However, for axial-vector couplings, the bosons couple to the total spin of muons, which is much smaller than their total charge. Further numerical calculations for axial-vector radiation would face uncertainties related to the neutron star's magnetic field and potential divergences as $m_{A^{\prime}} \rightarrow 0$, due to the nonconservation of the axial-vector current. Thus, we do not consider these constraints in this work.

For models with a chiral anomaly, or in anomaly-free models that include heavy hidden fermions, the decay $Z \rightarrow \gamma A^{\prime}$ can have a non-zero width. Such rare $Z$-boson decays place strong limits on these scenarios, typically requiring $g' \lesssim 10^{-13}\left(m_{A^{\prime}} / \mathrm{eV}\right) $ ~\cite{Dror:2017ehi, Dror:2017nsg}.
These bounds, however, are highly model-dependent and disappear entirely for the $L_\mu^{\rm AV}-L_\tau^{\rm AV}$ construction. By contrast, the limits we derive here depend only weakly on the ultraviolet completion of the theory and therefore remain robust across a broad class of models. 

\section{electron constraints}
\label{sec:electron}

Although our main focus is on the muon-coupled cases because they are directly related to the muon $(g-2)_\mu$ experiments, here we briefly discuss the electron-coupling case for completeness. We consider both the low energy effective model where only $Q_e = 1 \ne 0$ in Eq.~\eqref{eq:anomaly_free_model} which we denote it as $L^{\rm AV}_e$ model, together with the anomaly free models $L^{\rm AV}_e - L^{\rm AV}_\mu$ (only $Q_\tau = 0$) and $L^{\rm AV}_e - L^{\rm AV}_\tau$ (only $Q_\mu = 0$). For nonrelativistic electrons, the interaction Hamiltonian is
\begin{equation}
H^{\rm AV}_{e, \rm{int}} = g'_{e} \vec{A}' \cdot \vec{\sigma}_e, 
\end{equation}
where the vector field $\vec{A'}$ is given by Eq.~\eqref{eq:axial-vector-potential}, and $\vec{\sigma}_e$ is the electron spin, $g'_e$ denotes the coupling constant involving electron coupling. 

The most relevant experiments that measure the influence from an external field to electron spin is the  torsion pendulum by Heckel et al.~\cite{Heckel:2008hw}.
In such experiment, an energy shift is
\begin{align}
\Delta E  = - N_p \vec{\sigma}_p \cdot \vec{\beta}
\end{align}
causes a measurable torque $\vec{\tau} = N_p \vec{\sigma}_p \times \vec{\beta} $, where $\vec{\beta}$ represent an vector field, $ N_p$ is the net number of polarized spins, and $\vec{\sigma}_p $ represents the pendulum's spin orientation. The current torsion pendulum experiments have already provided strongest constraints~\cite{Cong:2024qly} (except some astrophysical bounds) to the vector-axial-vector (V-AV) model in low mass region, where a light boson has vector coupling to nucleon and AV coupling to electron. In such V-AV model, the potential is given by~\cite{Dobrescu:2006au, Fadeev:2018rfl}
\be
V_{12+13} = g_A^{e} g_V^N \frac{e^{-m_{V} r}}{4 \pi r} \vec{v}_{\rm rel} \cdot \vec{\sigma}_e,
\label{eq:v1213}
\ee
where $g_A^{e} g_V^N$ is the coupling constants of the model, $m_{V}$ is the vector boson mass, $r$ is the distance between nucleons (from Sun) and electrons (on Earth), and $\vec{v}_{\rm rel}$ is their relative velocity.

In the massless limit, for V-AV model, the potential from the Sun will be dominate due to the large amount of nucleons inside Sun~\cite{Heckel:2008hw}. Therefore, for both of the two model (axial-vector model and V-AV model), the revolution of $\vec{A}'$ and $\vec{v}_{\rm rel}$ will have the same period while only be different by a $\pi/2$ phase, due to $\vec{A}'$ and $\vec{v}_{\rm rel}$ being almost perpendicular to each other. Consequently, the energy shifts induced in the two models exhibit similar time-dependent behavior and can be constrained using the same experimental data.
The current constraints from torsion pendulum experiment to the V-AV model~\cite{Heckel:2008hw} is 
\begin{align}
g_A^{e} g_V^N < 1.2 \times 10^{-56} ~~(\rm  massless limit).
\end{align}
By requiring that the energy shift in our axial-vector model not exceed that corresponding to this experimental upper bound, we obtain
\begin{align}
    |g'_{e} \vec{A}'| <  1.2 \times 10^{-56} \left|\frac{N_n}{4 \pi R} \vec{v}_{\rm rel} \right|,
\end{align}
where $N_n$ is the total number of nucleons in Sun and $R$ is distance between Sun and Earth.
This will translate to a constraint on electron coupling
\begin{equation}
\begin{aligned}
&g'_{e} < 2.4 \times 10^{-22} ~(L^{\rm AV}_e ~\rm case),\\
&g'_{e} < 3.0 \times 10^{-22} ~ (L^{\rm AV}_e - L^{\rm AV}_\mu ~ \rm case),\\
&g'_{e} < 3.3 \times 10^{-22} ~ (L^{\rm AV}_e - L^{\rm AV}_\tau ~ \rm case).
\end{aligned}
\end{equation}

Even stronger limits arise from experiments probing electron–spin interactions within Earth’s geomagnetic field.
Using polarized electrons inside the Earth as an effective source, Ref.~\cite{Hunter:2013hza} derived
\begin{align}
    g'_{e} \lesssim 7 \times 10^{-24},
\end{align}
which is one or two order of magnitude stronger than our constraint.

\section{Conclusion}
\label{sec:concl}

We have investigated long-range axial--vector interactions that couple to leptons, focusing on the muon sector. Two realizations were analyzed: a minimal effective model only acting on muon flavor and an anomaly-free axial \(U(1)'\) model with inter-generation cancellations, $L_\mu^{AV} - L_\tau^{AV}$. In both cases, left-handed neutrinos source the new gauge field \(A'\). As a result, the solar neutrino flux generates a potential at Earth that can act on muon spins.

A central outcome of this setup is a distinctive diurnal modulation in storage-ring measurements of the muon anomalous magnetic moment $(g-2)_\mu$. The neutrino-sourced $A^{\prime}$ field produces a sign-changing contribution-positive during the day and negative at night-superimposed on a small time-independent offset. We find that the modulated component provides slightly stronger constraints than the unmodulated one. Based on a simple sensitivity estimate using the day-night asymmetry accumulated over a year of data, current $(g-2)_\mu$ measurements can already probe axial couplings as small as $g^{\prime} \lesssim \mathcal{O}\left(10^{-19}\right)$ in both model frameworks. A dedicated analysis of real data that fits the full time-dependent waveform could further improve these bounds.

For completeness, we also considered the corresponding effect on electron spins. By mapping the neutrino-sourced potential onto electron spin-sensor experiments, we obtained a complementary constraint of $g_e^{\prime} \lesssim \mathcal{O}\left(10^{-22}\right)$.

\section{Acknowledgments}
The work of J.L. is supported by the National Science Foundation of China under Grant No. 12235001, No. 12475103 and State Key Laboratory of Nuclear Physics and Technology under Grant No. NPT2025ZX11.
The work of X.P.W. is supported by National Science Foundation of China under Grant No. 12375095, and the Fundamental Research Funds for the Central Universities.
J.L. and X.P.W. thank APCTP, Pohang, Korea, for their hospitality during the focus program [APCTP-2025-F01], from which this work greatly benefited. J.L. and X.P.W. also
thank the Mainz Institute for Theoretical Physics (MITP) of the PRISMA+ Cluster of Excellence
(Project ID 390831469) for its hospitality and partial support during the completion of this work. The authors gratefully acknowledge the valuable discussions and insights provided by the members of the Collaboration of Precision Testing and New Physics.

\section*{Appendix: potential in high mass limit}
\label{App-1}

In this section, we discuss the potential in high mass limit, where we have $m_{A'} \gg R^{-1}$. In such limit, the contribution from neutrino with distance larger than $m^{-1}_{A'}$ will be highly depressed by the $e^{- m_{A'} \Delta r}$ term. Therefore, we choose Earth as the origin of the coordinate to compute the potential, where the position of Sun is chosen to be $(0, 0, R)$. Then the potential is given by
\begin{equation}
\begin{aligned}
A'^0 &= g' q^{\rm eff}_\nu \int d^3 x \frac{n_\nu R^2}{r^2 + R^2 - 2 r R \cos\theta} \frac{1}{4 \pi r} e^{- m_{A'} r} \\
& = g' q^{\rm eff}_\nu \int d^3 x \frac{n_\nu }{1 + \delta^2 - 2 \delta \cos\theta} \frac{1}{4 \pi r} e^{- m_{A'} r} 
\end{aligned}
\end{equation}
and 
\begin{equation}
\begin{aligned}
|\vec{A}'| &= g' q^{\rm eff}_\nu\int d^3 x \frac{n_\nu R^2 (R- r \cos\theta)}{\left(r^2 + R^2 - 2 r R \cos\theta\right)^{\frac{3}{2}}} \frac{1}{4 \pi r} e^{- m_{A'} r}\\
&= g' q^{\rm eff}_\nu \int d^3 x \frac{n_\nu ( 1 - \delta \cos\theta)}{\left(1 + \delta^2 - 2 \delta \cos\theta\right)^{\frac{3}{2}}} \frac{1}{4 \pi r} e^{- m_{A'} r}
\end{aligned}
\end{equation}
where $n_\nu$ is the number density of neutrinos, $r$ is the distance between neutrino and Earth, and $\delta = r/R$. In the high mass limit, we only need to consider the contribution from $\delta \ll 1$. Therefore, we can
perform the Taylor expansion on the term with $\delta$ and retain only up to the first-order terms. Then, we have
\begin{equation}
\begin{aligned}
A'^0 & \approx g' q^{\rm eff}_\nu \int r  d\cos\theta   d r\  \frac{n_\nu}{2} (1 + 2 \delta \cos\theta )  e^{- m_{A'} r} \\
& = g' q^{\rm eff}_\nu \frac{n^0_\nu}{m^2_{A'}}, 
\end{aligned}
\end{equation}
and similarly
\begin{equation}
\begin{aligned}
|\vec{A}'| & \approx g' q^{\rm eff}_\nu \int r  d\cos\theta  d r\  \frac{n_\nu}{2} (1 + 2 \delta \cos\theta )  e^{- m_{A'} r} \\
& = g' q^{\rm eff}_\nu \frac{n^0_\nu}{m^2_{A'}}. 
\end{aligned}
\end{equation}

We can find that, unlike the massless limit where $A'^0$ is one order of magnitude larger than $|\vec{A}'|$, in high mass limit $A'^0$ and $|\vec{A}'|$ are nearly the same. Numerical result also support this conclusion. As $R^{-1} \sim 10^{-18}$ eV, such high mass limit will hold when $m_{A'}$ is at least one order of magnitude larger than $R^{-1}$, in other word, $m_{A'} \gtrsim 10^{-17}$ eV. As shown in Fig.~\ref{fig:constraint}, numerical results are also consistent with this conclusion.

\bibliography{ref.bib}{}
\bibliographystyle{utphys28mod}
\end{document}